# Selective transverse mode emission from all polarization-maintaining fiber lasers

SHA WANG,[1,*] XINGWEN HU,[1] ZHICHENG ZHANG,[1] SHUTONG WANG[1,*] AND SHOUHUAN ZHOU[1,2]

[1] *College of Electronics and Information Engineering, Sichuan University, Chengdu 610064, China*
[2] *North China Research Institute of Electro-Optics, Beijing 100015, China*
*\*shawang@scu.edu.cn  wangst@scu.edu.cn*

**Abstract:** We demonstrated an all polarization-maintaining fiber (PMF) laser which could offer selective transverse mode emission. Offset and cross splicing between PMF sections was used to generate higher order mode and to yield a birefringence induced mode filter. Polarization rotation technique (PRT) was used to select the output beam profile. Both linear cavity and ring cavity were explored. Experimentally, 360° rotating $LP_{11}$ mode and OAM vortex beam with $\pm 1$ topological charges have been realized from an all PMF linear cavity laser and a ring cavity laser respectively.

## 1. Introduction

Selective transverse mode emission from a fiber laser can offer flexibility and can be used for different applications. Besides fundamental mode, higher-order modes, such as $LP_{11}$ mode and OAM (orbital angular momentum) vortex beams are also expected in many applications. For example, the waveguide dispersion for $LP_{11}$ mode can be used to compensate the chromatic dispersion in a fiber system, because that it can be tailored at the fiber cutoff wavelength [1]. Moreover, $LP_{11}$ modes are the basis for the generation of vector and OAM vortex beams in optical fibers [2], and it also has the ability to suppress the nonlinear distortion of signal spectrum in the power amplifier and is suitable for power scaling [3]. Especially orientation rotated $LP_{11}$ modes can be used for mode division multiplexing (MDM) technique. By aligning the orientation of the LP mode with the de-multiplexers, the MDM can be implemented with low mode cross talk [4]. OAM vortex beams, which have the characters of phase singularity and donut shaped beam profile, have been widely used in the field of optical tweezers [5], optical manipulation [6], optical capture [7], microscopic imaging [8], quantum information processing [9] and so on.

  Several approaches have been proposed to generate controlled transverse mode output from a fiber system. Specially designed fibers, such as spatially doped gain fibers [10] or refractive index managed vortex fibers [11] can be used to emit select proper mode from a fiber laser. However, these fibers are normally not off-the-shelf. Another possibility is to use offset launch [12], mode selective couplers [13] or long period gratings [14] to convert $LP_{01}$ mode to $LP_{11}$ mode. Proper mode filter methods, for instance utilizing fiber gratings can further increase the purity of the higher order mode from a fiber laser. Meanwhile, fiber lasers that directly oscillating at higher order $LP_{11}$ mode have also been reported to increase the lasing efficiency [15]. Most of them used few-mode fiber Bragg gratings [16, 17] or wavelength-division-multiplexing MSC [18] as mode filters. Among them, intra-cavity polarization filter, which is composed of a half wave plate and a polarizing beam splitter (PBS), was used by Lin [19] to select $LP_{01}$ mode and donut shaped $LP_{11}$ mode in a fiber laser cavity. It is found that the polarization state evolution along the fiber is mode dependent, because of the stress or bending induced birefringence in fibers. Hence, polarization rotation technique (PRT) in mode locking fiber lasers can provide us lessons for the mode selection in a fiber laser.

  Although most of the controlled mode outputs are realized in non-polarization maintaining fibers, there are still some papers which reported higher order mode generated from polarization maintaining (PM) fibers. Normally, using PMFs increases the environmental stability performance of a laser. The mode controlling can be realized both in a passive generator (converting the fundamental Gaussian beams into other modes outside a laser) and in an active generator (direct generating higher order modes from a laser cavity). For example, in the passive way, Hong [20] reported a tunable mode rotator based on a piece of few mode PMF with a free structure. The input polarized $LP_{11}$ mode was misaligned by an angle α regards to the slow axis. The output $LP_{11}$ mode was rotated by an angle of 2α if the wavelength was tuned to get π phase difference between even and odd $LP_{11}$ modes. Zhang [21] used a long-period fiber grating based on a two-mode PM photonic crystal fiber to convert $LP_{01}$-$LP_{11}$ mode. Gopinath's group [22, 23] reported generation of OAM modes in a piece of PMF. Two offset launch spots at different position of the PMF were used to excite even and odd $LP_{11}$ modes at the same polarization state. The relative phase between two $LP_{11}$ modes were adjusted using a piezo-driven delay stage in the free space. Similar experiments were performed by the same group later to achieve $HG_{11}$ mode and the second order OAMs [24]. Zeng [2] reported a fiber OAM generator based on a piece of few mode PMF. $LP_{11}$ mode was incident to a piece of FM-PMF and aligned at a specified angle to the slow axis. Thanks to the birefringence of the PMF, ±1 OAM can be obtained by properly tune the laser wavelength. Dong [25] reported stable OAM mode generation by combining two linearly polarized $LP_{11}$ modes output from two PMF lasers using PM-LPFGs. The phase difference between these two $LP_{11}$ modes was controlled by a fiber optic stretcher. In the active way, Zhao [26] obtained cylindrical vector beam generation by using vortex phase plate inside an all-PM Er-doped fiber laser cavity. Zhang [27] made a mode converter by using a piece of polarization-maintaining fiber and a ring core fiber based long period fiber grating and used it in a fiber laser. In their paper, the PMF was used to maintain and control the polarization direction of the input $HE_{11}$ mode into the ring core fiber. Stable $TE_{01}$ and $TM_{01}$ mode were obtained. To the best of our knowledge, direct generation of OAM beams from a PMF laser has not been reported yet.

In this paper, we developed a new structure using polarization rotation technique (PRT) to select the output beam profile directly from an all PMF laser. Offset and cross splicing was used to generate higher order mode in a few-mode PMF and to yield a birefringence induced mode filter. A linear laser cavity was firstly built. Stable 360 °rotating $LP_{11}$ mode was realized in the experiments. Owing to the large birefringence of the PMF, the crosstalk between $LP_{01}$ and $LP_{11}$ mode can be significantly reduced. The generation of rotating $LP_{11}$ mode was robust and against large perturbations. A PMF ring laser cavity was also built, and OAM vortex beam output with ±1 topological charge was achieved directly from this PMF ring fiber laser.

## 2. Experimental setup

There are two steps to control the output mode in an all PMF laser using PRT. The first step is to select the higher order $LP_{11}$ mode as output. If all PMFs are spliced with the matched slow axes, there will be no polarization rotating phenomenon. However, if a splicing angle is intentionally introduced between two PMFs, we will have light components along both fast and slow axes of the PMF. After passing through the PMF, the light will become elliptically polarized. For a PMF, the beating length for $LP_{01}$ mode and $LP_{11}$ mode are different. Hence, after propagating a certain distance in the fiber, the $LP_{01}$ mode and the $LP_{11}$ mode have different elliptical polarization states. With an additional QWP and an HWP, the polarization of the $LP_{01}$ mode and $LP_{11}$ mode can be tuned to S and P polarization respectively. Hence, a polarization beam splitter (PBS) can be used to filter out the $LP_{11}$ mode from the laser cavity. For PM fibers, $LP_{11}$ mode is Eigenmode of the fiber. The output $LP_{11}$ mode is aligned with an angle to the slow axis. Therefore, if the slow axis of the FM-PMF output face is rotated, the output $LP_{11}$ mode orientation is consistently changed. The second step is to tune the phase difference between the $LP_{11}$ even and odd modes. If the $LP_{11}$ mode axis and the fiber slow axis is aligned with an angle α at the input of the FM-PMF, and the laser polarization and the fiber slow axis is aligned with an angle of β. The $LP_{11}$ mode then can be decomposed into two orthogonal degenerate modes $LP_{11,even}$ and $LP_{11,odd}$ along slow and fast axes. After propagating through a piece of FM-PMF, the phase difference between $LP_{11,even}$ and $LP_{11,odd}$ will be $\Delta\Phi=\varphi_2-\varphi_1=2\pi L\cos(\beta)\Delta n_{eff,x}/\lambda-2\pi L\sin(\beta)\Delta n_{eff,y}/\lambda$, where $\Delta n_{eff,x}$ and $\Delta n_{eff,y}$ are the refractive index difference between the two orthogonal degenerated modes in the x and y directions, L is the fiber length, and λ is the laser wavelength. The phase difference can be tuned by changing the fiber length L, the laser wavelength λ or the input light polarization. In order to precisely tune the phase difference, ring cavity with an additional HWP to control the polarization in order to tune the phase difference ΔΦ is preferred. If ΔΦ is set as π/2, the $LP_{11}$ mode will be turned to OAM vortex beam.

We firstly built an all PMF linear cavity laser, the setup of which is shown in Fig.1. All fibers used in the cavity were Panda-type PMFs. A piece of 70 cm long Yb doped PMF (Liekki Yb300-6/125) was used as the gain medium. A single mode PMF coupled laser diode (LD) at 976 nm was used as pump source, and coupled into the cavity through a polarization maintaining fused wavelength division multiplexer (PMWDM). A 1×2 PMF coupler with a coupling ratio of 50:50 was used as one end mirror. PM1550 fiber was used as a two-mode PMF at 1030 nm. OSS (Offset Splicing) between PM980 fiber and PM1550 fiber was employed to excite high order mode excitation in the fiber laser. In the meanwhile, the PM980 fiber and PM1550 fiber were also cross spliced to yield a birefringence induced mode filter. A quarter wave plate (QWP) and a half wave plate (HWP) were used after the collimator to fine tune the polarization of $LP_{01}$ mode and $LP_{11}$ mode. The selected transverse mode laser output was ejected out directly from the polarization beam splitter (PBS). The collimator can be rotated 360 ° in order to achieve the rotated $LP_{11}$ mode.

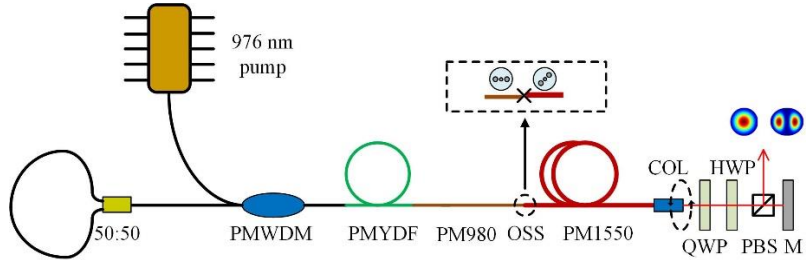

Fig. 1. Experimental setup of an all PMF laser cavity. PMWDM: polarization maintaining fused wavelength division multiplexer, PMYDF: Yb doped PMF (Liekki Yb300-6/125), COL: collimator, QWP: quarter wave plate, HWP: half wave plate, PBS: polarization beam splitter, M: silver mirror, OSS: offset splicing.

## 3. Experimental results

Figure 2 shows the experimental results of 360 ° rotating $LP_{11}$ mode. Firstly, $LP_{11}$ mode was achieved by rotating the QWP and HWP. Then the lobe orientation rotation was obtained by rotating the collimator (COL) shown in Fig. 1. The power almost keeps off the center of all the captured field from Fig. 1, which indicates that no mode coupling between $LP_{01}$ mode and the $LP_{11}$ mode was observed. The intensity of two lobes are slightly different, which should come from the residual $LP_{01}$ mode with extremely low power.

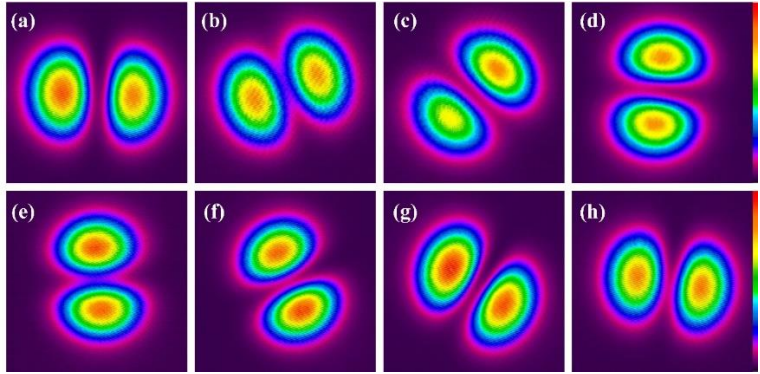

Fig. 2. Measured beam profile of 360 °rotating $LP_{11}$ mode.

The FM-PMF was bended on purpose in order to check the environmental stability of the generated $LP_{11}$ mode. Fig. 3 (a) shows the output beam profile when the bending diameter was 13 mm, while Fig. 3 (b) shows the output beam profile when the bending diameter was 5 mm. By changing the bending diameter, the lobe profile and orientation stayed the same, while the intensity of the two lobes changed a little bit. The intensity change is due to the modal interference of the low power $LP_{01}$ and the $LP_{11}$ modes.

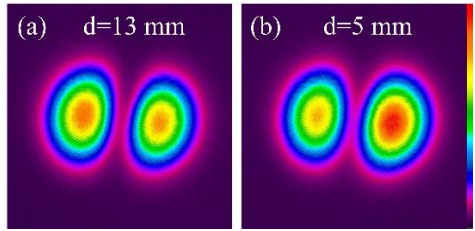

Fig. 3. Measured beam profile of $LP_{11}$ mode while the fiber is bended: (a) bending diameter is 13 mm; (b) bending diameter is 5 mm.

In order to achieve high purity vortex beam, ring cavity was employed. Experimental setup is shown in Fig. 4 (a). Offset and cross splicing was also employed between PM980 and PM1550 fibers. The beam came out from and coupled into the fibers via a fiber collimators (COL). Bulk waveplates HW1, HW2 and QW are used to manipulate the polarization of $LP_{01}$ mode and $LP_{11}$ mode. Compared with linear cavity, laser polarization of the input beam to PM1550 can be precisely controlled by HW2, which makes it easier to achieve high purity OAM vortex beams. A Mach–Zehnder interferometer was employed to discover the chirality of the output donut beam. The beam profiles and the interference patterns are shown in Fig. 4 (b)-(e). Both $OAM_{+1}$ and $OAM_{-1}$ modes were obtained experimentally. Generation of the vortex beams was not as robust as the $LP_{11}$ modes, i.e. bending the fiber will influence the vortex beam profiles. The reason might be that precise $\pi/2$ phase difference between even and odd $LP_{01}$ modes is required to generate a vortex beam. Although large perturbation of the PM fibers will not change the birefringence of the fiber, the effective length will be influenced, i.e. the phase difference will be changed. However, after beam profile distortion caused by perturbation of the fibers, the donut shaped beam can always be got back by slightly adjusting the waveplates. The output vortex beam is a linearly polarized OAM mode, of which the purity can be calculated by analysis of the amplitude and phase spectrum of the data ring derived from the electric field intensity [28]. Fig. 5 (a) shows the azimuthal intensity profile of the $OAM_{+1}$ vortex beam for radius $r_0$ taken from the gray scale image. Fig. 5 (b) and (c) shows the Fourier series analysis of the data ring of $OAM_{+1}$. The mode purity is calculated as 91.1% for generated $OAM_{+1}$ mode from the Fourier series coefficients. Fig. 5 (d) shows the azimuthal intensity profile of the OAM$_{-1}$ vortex beam for radius r$_0$ taken from the gray scale image. Fig. 5 (e) and (f) shows the Fourier series analysis of the data ring of OAM$_{-1}$. The mode purity is calculated as 96.55% for generated OAM$_{-1}$ mode. We would like to mention that besides OAM beams, rotating LP$_{11}$ mode can also be achieved from the ring cavity. From the analysis, the mode impurity mainly come from the residual LP$_{01}$ mode.

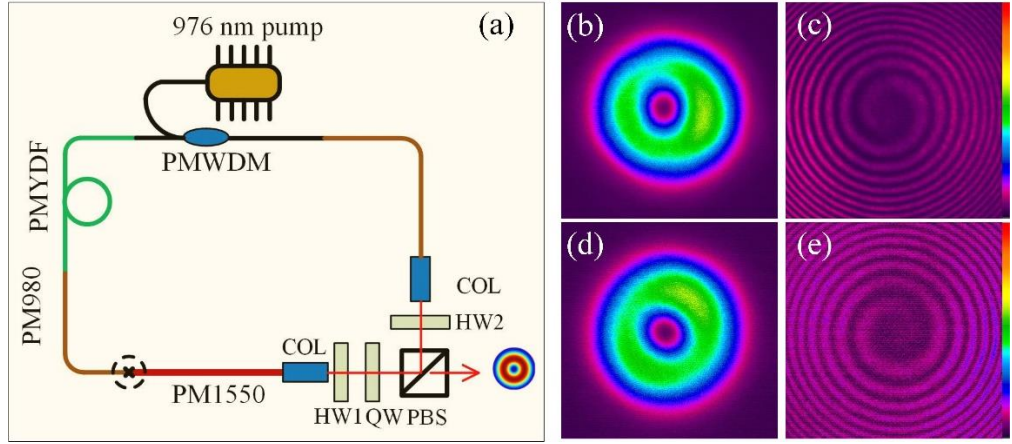

Fig. 4. (a) Experimental setup of the PM fiber ring cavity laser. PMWDM: polarization maintaining fused wavelength division multiplexer, PMYDF: Yb doped PMF (Liekki Yb300-6/125), COL: collimator, QWP: quarter wave plate, HWP: half wave plate, PBS: polarization beam splitter. X indicates the offset and cross splicing point; (b) Measured intensity profile of the OAM$_{+1}$ vortex beam; (c) the interference pattern of clockwise spiral form; (d) Measured intensity profile of the OAM$_{-1}$ vortex beam; (e) the interference pattern of anticlockwise spiral form.

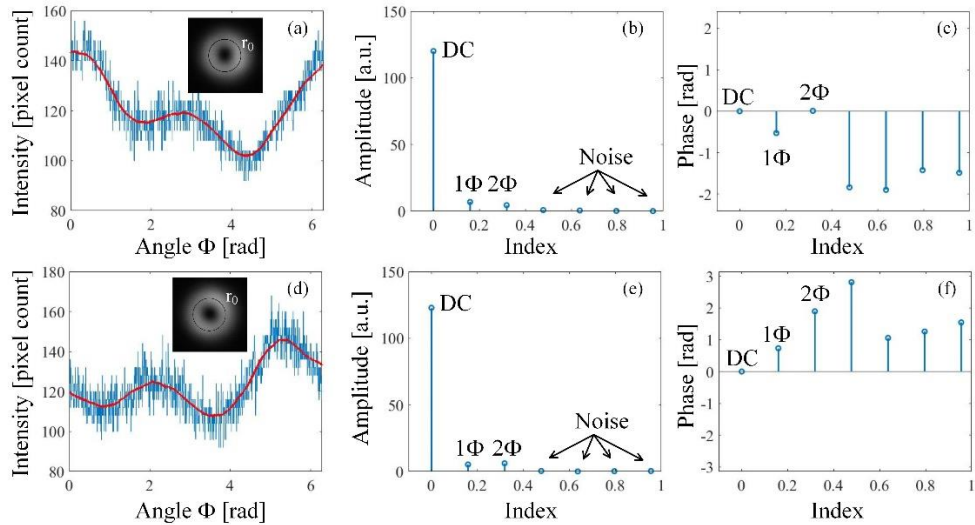

Fig. 5. Azimuthal intensity profile of the OAM$_{+1}$ beam for radius r$_0$; (b) amplitude spectrum of data ring; (c) phase spectrum of data ring; Azimuthal intensity profile of the OAM$_{-1}$ beam for radius r$_0$; (b) amplitude spectrum of data ring; (c) phase spectrum of data ring.

## 4. Conclusion

In conclusion, we have demonstrated the generation of $LP_{11}$ mode and linear polarized OAM vortex beams from all PMF lasers. The output $LP_{11}$ mode can be rotated 360 ° and is stable against bending of the fibers. OAM vortex beams with ±1 topological charges can be obtained from a PMF ring cavity and the mode purity is better than 90%. The laser setup is simple and compact. A PMF polarization splitter and precisely cross splicing between PMF sections can be used to replace the bulk PBS and waveplates to make an all fiber structure. In the future, how to further improve the mode purity and laser stability of OAM vortex beams should be studied.

**Funding**

National Natural Science Foundation of China (NSFC) (61975137)

**Disclosures**

The authors declare that there are no conflicts of interest related to this article.